\documentclass[a4paper]{article}

\usepackage{INTERSPEECH2020}
\usepackage[labelformat=simple]{subcaption}
\usepackage{color}
\usepackage{ulem}

\title{Depthwise Separable Convolutional ResNet with Squeeze-and-Excitation Blocks for Small-footprint Keyword Spotting}
\name{Menglong Xu, Xiao-Lei Zhang}
\address{
  CIAIC, School of Marine Science and Technology,
   Northwestern Polytechnical University, China}

\email{mlxu@mail.nwpu.edu.cn, xiaolei.zhang@nwpu.edu.cn}

\begin{document}

\maketitle

\begin{abstract}
  One difficult problem of keyword spotting is how to miniaturize its memory footprint while maintain a high precision.
  Although convolutional neural networks have shown to be effective to the small-footprint keyword spotting problem, they still need hundreds of thousands of parameters to achieve good performance.
  In this paper, we propose an efficient model based on depthwise separable convolution layers and squeeze-and-excitation blocks. Specifically,
  we replace the standard convolution by the depthwise separable convolution, which reduces the number of the parameters of the standard convolution without significant performance degradation. We further improve the performance of the depthwise separable convolution by reweighting the output feature maps of the first convolution layer
  with a so-called squeeze-and-excitation block. We compared the proposed method with five representative models on two experimental settings of the Google Speech Commands dataset. Experimental results show that the proposed method achieves the state-of-the-art performance. For example, it achieves a classification error rate of 3.29\% with a number of parameters of 72K in the first experiment, which significantly outperforms the comparison methods given a similar model size. It achieves an error rate of 3.97\% with a number of parameters of 10K, which is also slightly better than the state-of-the-art comparison method given a similar model size.
\end{abstract}

\section{Introduction}

    Keyword spotting (KWS) aims at detecting predefined keywords in an audio stream.
    A common approach for KWS is based on large vocabulary continuous speech recognition \cite{miller2007rapid,bai2016end}. It costs huge memory footprint and has a high latency, so that it is often used for the keyword search in large databases.
    Another approach is based on keyword/filler hidden Markov models (HMMs) \cite{silaghi2005spotting}, which requires a high computational cost and is therefore difficult to be applied to on-device applications. This paper focuses on small-footprint KWS, which requires small memory footprint and low computational power. This kind of technology is able to run on low-resource devices. It provides a fully hands-free way for users to control intelligent devices.

    Recently, deep neural network (DNN) based approaches yield significant improvement over the conventional methods in small-footprint KWS. DeepKWS \cite{chen2014small} regards keyword spotting as a classification problem and trains a DNN to directly predict the subword units of keywords. It achieves significant improvement over the HMM-based methods, e.g. \cite{silaghi2005spotting}, with small footprint and low computational cost. Because DNN does not consider the local temporal and spectral correlation of speech, Sainath and Parada \cite{sainath2015convolutional} proposed to replace DNN by convolutional neural network (CNN), which results in better performance with smaller memory footprint than DNN. However, the size of the receptive field of CNN is usually limited, which cannot grasp enough temporal correlation of speech. To overcome this problem,
    Tang and Lin \cite{tang2018deep} proposed a residual network (ResNet) based KWS system where they used dilated convolution to enlarge the size of the receptive field exponentially with the depth of the network.
    However, the ResNet based method still needs several hundreds of thousands of parameters to achieve the state-of-the-art performance. To further reduce the memory footprint, a number of recent works applied time delay neural network (TDNN), attention mechanism, and temporal convolutional network (TCN) to KWS, see e.g. \cite{sun2017compressed,shan2018attention,choi2019temporal}.
    In \cite{zhang2017hello}, Zhang \textit{et al.} adapted MobileNet \cite{howard2017mobilenets} that was originally designed for image classification to KWS, where MobileNet reduces the number of parameters and computational costs by a so-called depthwise separable convolution structure \cite{sifre2014rigid}. However, if the method uses numerous ReLU activation functions after the convolution operations, the representational ability of the model may be hurt \cite{sandler2018mobilenetv2}, and moreover, the method adopts a conventional convolution structure, which is inefficient in propagating gradients across layers. To summarize, although a number of new architectures have been proposed, they still need a lot of parameters, which does not fully meet the requirement of modern low-resource devices.

    Motivated by \cite{tang2018deep,zhang2017hello,sandler2018mobilenetv2}, we propose the \textit{depthwise separable convolution based ResNet} (DS-ResNet), which is a stack of depthwise separable convolution layers with residual connections. This structure not only improves the representation ability over \cite{zhang2017hello} but also results in smaller memory footprint than \cite{tang2018deep}.
    To further improve the performance of the proposed method, we add a squeeze-and-excitation block \cite{hu2018squeeze} over the output of the bottom convolutional layer of DS-ResNet. We compared DS-ResNet with ResNet \cite{tang2018deep}, TC-ResNet \cite{choi2019temporal}, DS-CNN \cite{zhang2017hello}, DenseNet-BiLSTM \cite{ZengEffective} and tdnn-swsa \cite{bai2019time}. Experimental results on the Google speech commands dataset demonstrate that DS-ResNet outperforms the comparison methods in terms of classification errors with fewer parameters than the latter.

    The remainder of the paper is organized as follows.
    Section \ref{sec:2} introduces the proposed DS-ResNet. Section \ref{sec:3} presents the experimental setup and results.
    Section \ref{sec:4} concludes the paper.

\section{Algorithm description}\label{sec:2}

 \subsection{Network structure}

 \begin{figure}[t]
     \centering
     \includegraphics[scale=0.39]{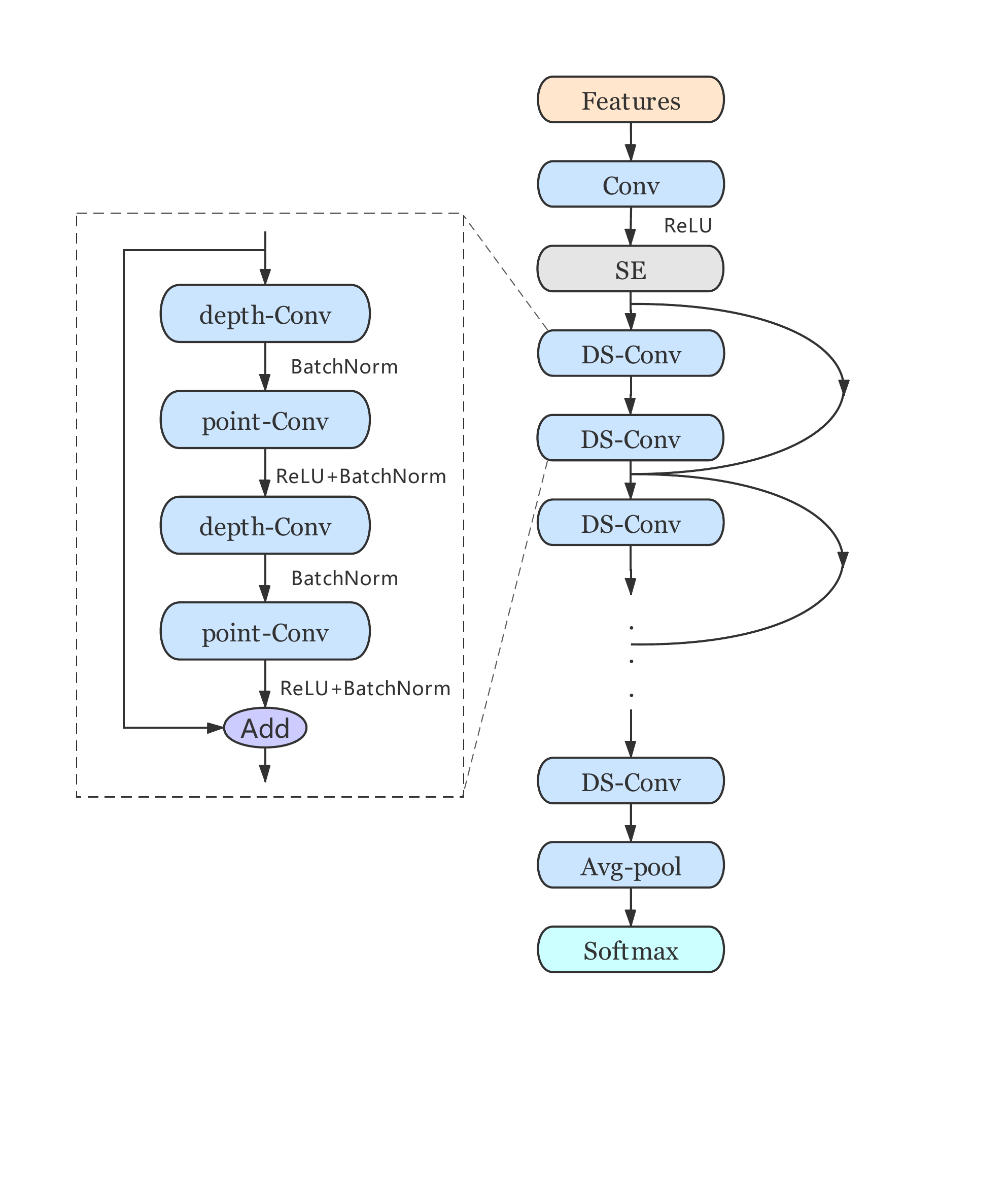}
     \caption{Model architecture of the proposed DS-ResNet. The content in the rectangle module describes a magnified residual block of two depthwise separable convolution layers.}
     \label{fig:model}
    \end{figure}

     \begin{figure*}[htbp]
      \begin{subfigure}{.33\textwidth}
        \setcaptionwidth{4.3cm}
        \centering
        \includegraphics[width=5cm]{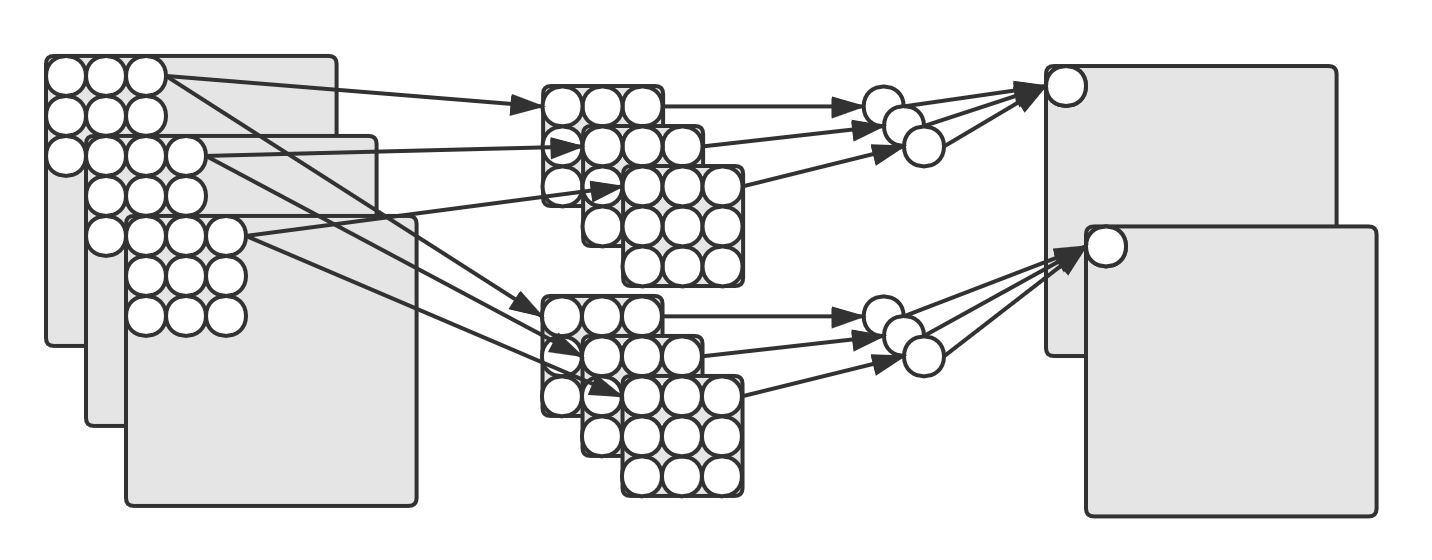}
        \caption{Standard convolution.}
        \label{fig:dsc_a}
      \end{subfigure}
      \begin{subfigure}{.33\textwidth}
        \setcaptionwidth{4.3cm}
        \centering
        \includegraphics[width=5cm]{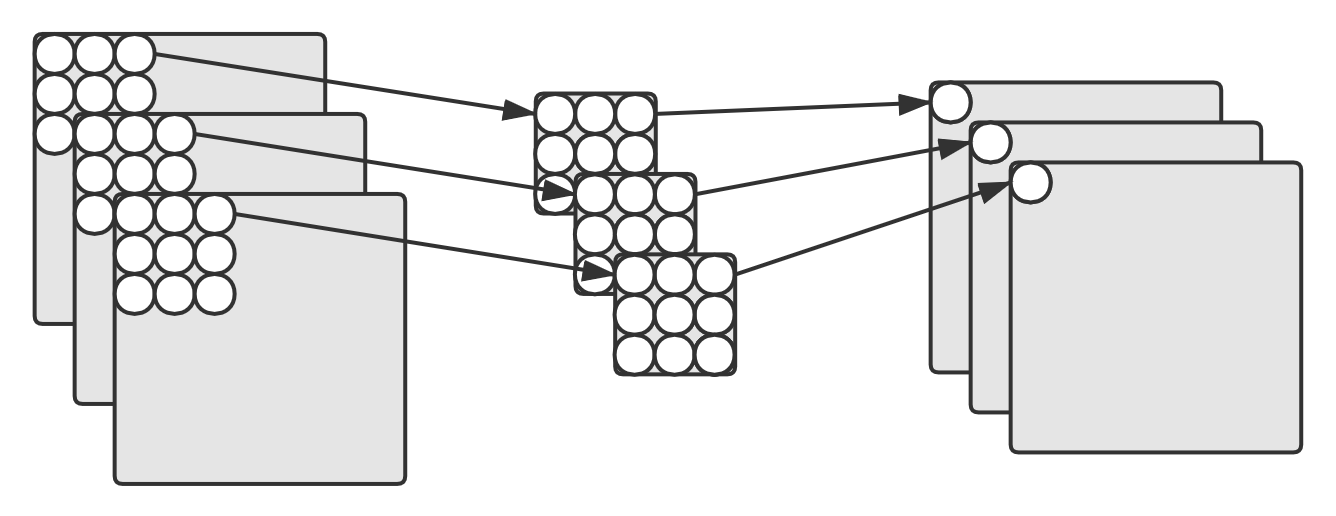}
        \caption{Depthwise convolution.}
        \label{fig:dsc_b}
      \end{subfigure}
      \begin{subfigure}{.33\textwidth}
        \setcaptionwidth{4.3cm}
        \centering
        \includegraphics[width=5cm]{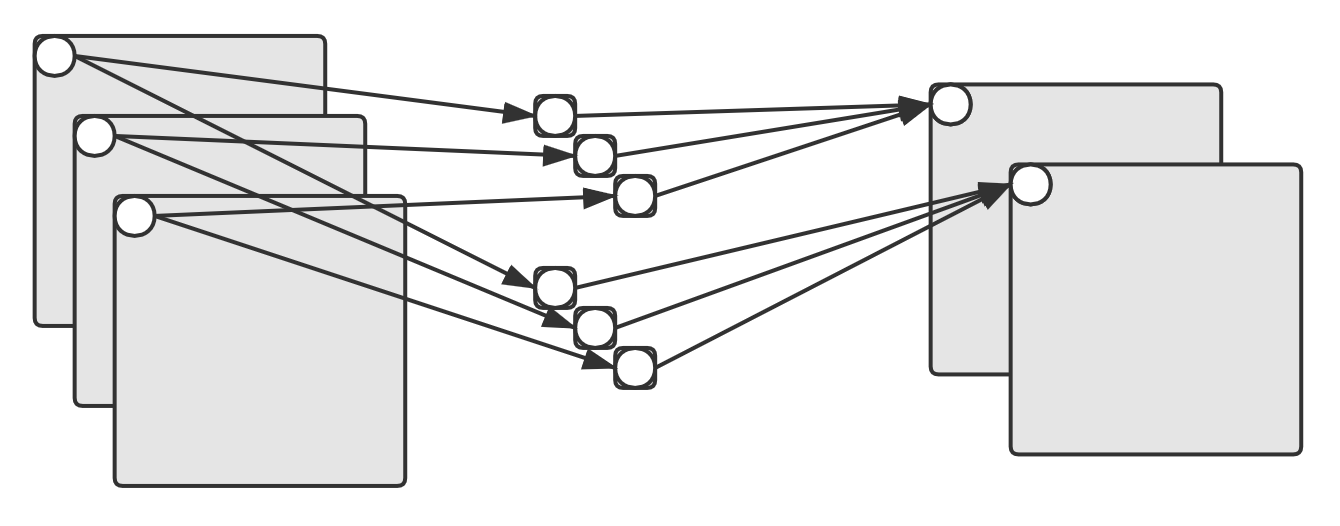}
        \caption{Pointwise convolution.}
        \label{fig:dsc_c}
      \end{subfigure}
      \caption{Comparison of three variants of the convolution operation, where we show $3$ input channels, $2$ output channels, and a kernel size of $3\times3$.)}
      \label{fig:dsc}
    \end{figure*}

    As shown in Figure \ref{fig:model}, our entire architecture starts with a standard bias-free convolution layer (Conv) with weight $\textbf{W}\in\mathbb{R}^{m\times r\times n}$, where $m$ and $r$ are the height and width of the convolution kernel respectively, and $n$ is the number of filters (i.e., the number of the output channels). The model takes the output of the first convolution layer as the input of a squeeze-and-excitation layer (SE) which is used to reweight the output feature maps.
     Then, the output of the squeeze-and-excitation layer is the input of a chain of residual blocks, followed by a separate non-residual depthwise separable convolution layer (DS-Conv) which consists of a depthwise convolution layer (depth-Conv) and a pointwise convolution layer (point-Conv).
    Finally, the output of the model is composed of an average-pooling layer (Avg-pool) followed by a fully-connected softmax layer (Softmax).
    Additionally, a $(d_{w},d_{h})$ convolution dilation was used to increase the receptive field of the depthwise separable convolution layers.

 \subsection{Depthwise separable convolution}

     Depthwise separable convolution considers the channel realm and space realm separately. It factorizes a standard convolution into two simplified steps. The first step is a spatial feature learning step, named depthwise convolution. The second step is a channel combination step, named pointwise convolution. The most attractive property of the depthwise separable convolution is its low computational cost and small amount of parameters.

    Before describing the depthwise separable convolution, we first take a look at the computational cost of a standard convolution. As shown in Figure \ref{fig:dsc_a},
     given a $C_{in}\times H_{in}\times W_{in}$ input feature maps of a certain convolution layer where $C_{in}$ is the number of the input channels, and $H_{in}$ and $W_{in}$ are the spatial height and width of the input feature maps, a standard convolution operation operates over a joint ``space-cross-channels realm'' which
     applies $C_{out}$ filters of size $D_{K} \times D_{K} \times C_{in}$ to compute the output feature maps, where $D_{K}$ is the spatial dimension of the convolution filters, $C_{in}$ is number of the input channels, and $C_{out}$ is the predefined number of the filters (i.e.,
     channels of the output feature maps).
      The computational cost $\mathcal{C}$ and amount of parameters $\mathcal{S}$ of the standard convolution are:
     \begin{equation}
        \mathcal{C}_{\textit{(Conv)}}=C_{in}\times D_{K}\times D_{K}\times H_{in}\times W_{in}\times C_{out}
     \end{equation}
     \begin{equation}
        \mathcal{S}_{\textit{(Conv)}}=C_{in}\times D_{K}\times D_{K}\times C_{out}
     \end{equation}
     where we assume that the stride is 1, padding mode is set to ``same'', and the size of the output feature maps is $C_{out}\times H_{out}\times W_{out}$.

   \begin{figure}[t]
     \centering
     \includegraphics[width=8cm]{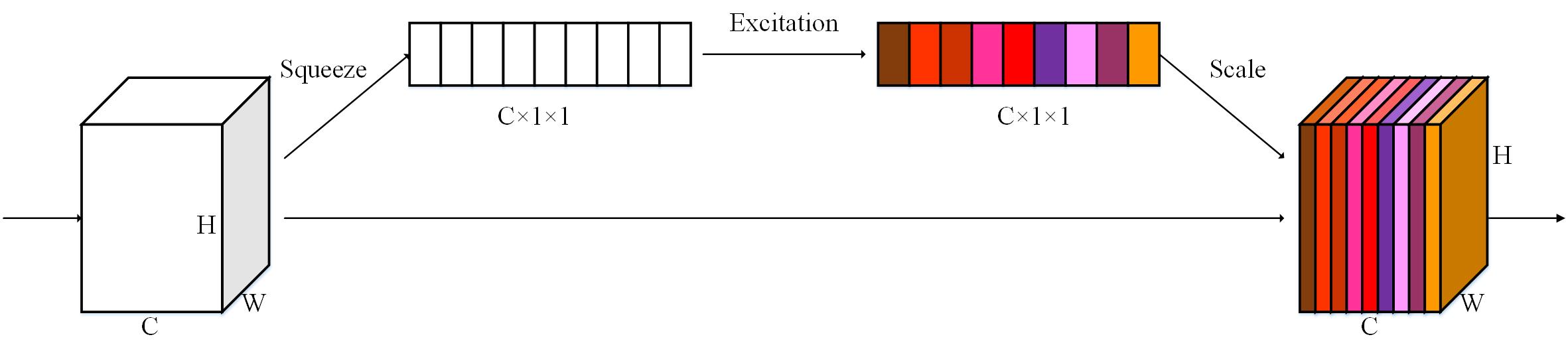}
     \caption{The architecture of the squeeze-and-excitation block.}
     \label{fig:se}
    \end{figure}

     Different from the standard convolution which filters and combines the input feature in one step, the filtering and combination step in the depthwise separable convolution is split into two successive steps. First, the depthwise convolution applies a single filter to each input channel (see Figure \ref{fig:dsc_b}). Then, the pointwise convolution applies a $1\times1$ convolution to combine the outputs of the depthwise convolution (see Figure \ref{fig:dsc_c}). The computational cost $\mathcal{C}$ and amount of parameters $\mathcal{S}$ of the depthwise convolution and pointwise convolution are
     \begin{equation}
        \mathcal{C}_{\textit{(depth-Conv)}}=1\times D_{K}\times D_{K}\times H_{in}\times W_{in}\times C_{in}
     \end{equation}
     \begin{equation}
        \mathcal{S}_{\textit{(depth-Conv)}}=1\times D_{K}\times D_{K}\times C_{in}
     \end{equation}
     and
     \begin{equation}
        \mathcal{C}_{\textit{(point-Conv)}}=C_{in}\times 1\times 1\times H_{in}\times W_{in}\times C_{out}
     \end{equation}
     \begin{equation}
        \mathcal{S}_{\textit{(point-Conv)}}=C_{in}\times 1\times 1\times C_{out}
     \end{equation}
     respectively, where we have made the same assumption as the standard convolution.

     To show the advantage of the depthwise separable convolution over the standard convolution apparently, we give an example as follows. When we set $C_{out}=64$ and $D_{K}=3$, the computational cost and model size of the depthwise separable convolution are only about 1/8 of those of the standard convolution.

 \subsection{Squeeze-and-excitation block}

    Squeeze-and-excitation block is a new architecture that aims to recalibrate the channel-wise feature responses adaptively by modeling the interdependency between the channels \cite{hu2018squeeze}.
    As illustrated in Figure \ref{fig:se},
    it consists of two successive operations---squeeze and excitation.
    The squeeze operation compresses the feature maps along the spatial dimension which extracts the mean of the feature maps for each channel.
    The excitation operation models the correlation between the channels, and then generates a weight for each channel.
    Finally, the output of the squeeze-and-excitation block is generated by multiplying the input feature maps of the block with the output weights of the excitation operation.

    In our implementation, the squeeze operation is an
    average-pooling layer. The excitation operation consists of two fully-connected layers that take the rectified linear units and sigmoid activation units as the hidden units respectively. The dimension between the two fully-connected layers can be adjusted by a hyperparameter $\alpha$. We set $\alpha=2^{-4}$ in this paper as \cite{hu2018squeeze} did.

\subsection{Model implementation}

    \begin{table}[tp]
      \caption{Parameter setting of \texttt{DS-ResNet18}, along with the number of parameters and multiplies.}
      \label{table:par_best}
      \centering
      \scalebox{0.83}{
      \begin{tabular}{c | c c c c c | c c}
        \hline
         &
        \multicolumn{1}{c}{$m$} &
        \multicolumn{1}{c}{$r$} &
        \multicolumn{1}{c}{$n$} &
        \multicolumn{1}{c}{$d_{w}$} &
        $d_{h}$ &
        \multicolumn{1}{c}{\#Parameters} &
        \multicolumn{1}{c}{\#Multiplies} \\
        \hline
        Conv & 3 & 3 & 64 & 1 & 1 & 576 & 2.3M
        \\
        SE & - & - & 64 & - & - & 512 & 576
        \\
        Res$\times$7 & 3 & 3 & 64 &
        $2^{\lfloor\frac{i}{3}\rfloor}$ & $2^{\lfloor\frac{i}{3}\rfloor}$ & 65.4K & 264M
        \\
        DS-Conv & 3 & 3 & 64 & 16 & 16 & 4672 & 18.9M              \\
        Avg-Pool & - & - & 64 & - & - & - & 64               \\
        Softmax & - & - & 12 & - & - & 768 & 768              \\
        \hline\hline
        Total  & - & - & - & - & - & 72K & 285M
        \\
        \hline
      \end{tabular}
      }
    \end{table}

    In this subsection, we configure the models for the experiment. The first model, named \texttt{DS-ResNet18}, achieves the highest accuracy with a small model size. It consists of 7 residual blocks, each of which contains 2 depthwise separable convolution layers and 64 input and output channels. Because there is an independent depthwise separable convolution layer before the average-pooling layer, the total number of the depthwise separable convolution layers is 15. The dilation in the $i$th layer was set to $2^{\lfloor\frac{i}{3}\rfloor}$. \texttt{DS-ResNet18} has roughly 72K parameters. It needs 285M multiplication operations to generate an output from an input time-frequency spectrum.
    The details of the model are listed in Table \ref{table:par_best}.

    To reduce the model footprint, the most efficient way is to use fewer filters in each convolution layer. Here we reduced the number of the input and output channels to $n=32$ for each depthwise separable convolution layer. To further reduce the model footprint, we further reduced the number of the depthwise separable convolution layers to 11. When the the number of the depthwise separable convolution layers is less than 13, the receptive field of the entire network cannot cover the entire input.
    To cover more context in the input,
    we added a $(2\times2)$ average-pooling layer after the squeeze-and-excitation layer.
    The model, named \texttt{DS-ResNet14}, has roughly 15.2K parameters, and needs 15.7M multiplies to generate an output.
    The details of the model are listed in Table \ref{table:par_mid}.

    The smallest model we have implemented contains 8 convolution layers. When the number of the convolution layer is less than 10, the residual connections seem unnecessary anymore. Therefore, we removed the residual connections. To keep the receptive field wide enough, we added a $(4\times2)$ average-pooling layer after the squeeze-and-excitation layer. This compact model, named \texttt{DS-ResNet10} (Note that it is no longer ResNet, but we followed the same naming convention), consists of 7 separable convolution layers without residual connections. It has about 10K parameters, and needs 5.8M multiplies to generate an output.
    The details of the model are listed in Table \ref{table:par_small}.

   \begin{table}[tp]
      \caption{Parameter setting of \texttt{DS-ResNet14}.}
      \label{table:par_mid}
      \centering
      \scalebox{0.83}{
      \begin{tabular}{c | c c c c c | c c}
        \hline
         &
        \multicolumn{1}{c}{$m$} &
        \multicolumn{1}{c}{$r$} &
        \multicolumn{1}{c}{$n$} &
        \multicolumn{1}{c}{$d_{w}$} &
        $d_{h}$ &
        \multicolumn{1}{c}{\#Parameters} &
        \multicolumn{1}{c}{\#Multiplies} \\
        \hline
        Conv & 3 & 3 & 32 & 1 & 1 & 288 & 1.2M
        \\
        SE & - & - & 32 & - & - & 128 & 160
        \\
        Avg-Pool & 2 & 2 & 32 & - & - & - & 32K               \\
        Res$\times$5 & 3 & 3 & 32 &
        $2^{\lfloor\frac{i}{3}\rfloor}$ & $2^{\lfloor\frac{i}{3}\rfloor}$ & 13.1K & 13.1M
        \\
        DS-Conv & 3 & 3 & 32 & 8 & 8 & 1312 & 1.3M              \\
        Avg-Pool & - & - & 32 & - & - & - & 32               \\
        Softmax & - & - & 12 & - & - & 384 & 384              \\
        \hline\hline
        Total  & - & - & - & - & - & 15.2K & 15.7M
        \\
        \hline
      \end{tabular}
      }
   \end{table}
   \begin{table}[t]
      \caption{Parameter setting of \texttt{DS-ResNet10}.}
      \label{table:par_small}
      \centering
      \scalebox{0.83}{
      \begin{tabular}{c | c c c c c | c c}
        \hline
         &
        \multicolumn{1}{c}{$m$} &
        \multicolumn{1}{c}{$r$} &
        \multicolumn{1}{c}{$n$} &
        \multicolumn{1}{c}{$d_{w}$} &
        $d_{h}$ &
        \multicolumn{1}{c}{\#Parameters} &
        \multicolumn{1}{c}{\#Multiplies} \\
        \hline
        Conv & 3 & 3 & 32 & 1 & 1 & 288 & 1.2M
        \\
        SE & - & - & 32 & - & - & 128 & 160
        \\
        Avg-Pool & 4 & 2 & 32 & - & - & - & 16K              \\
        DS-Conv$\times$7 & 3 & 3 & 32 &
        $2^{\lfloor\frac{i}{3}\rfloor}$ & $2^{\lfloor\frac{i}{3}\rfloor}$ & 9.2K & 4.6M
        \\
        Avg-Pool & - & - & 32 & - & - & - & 32               \\
        Softmax & - & - & 12 & - & - & 384 & 384              \\
        \hline\hline
        Total  & - & - & - & - & - & 10K & 5.8M
        \\
        \hline
      \end{tabular}
      }
   \end{table}

\section{Experiments}\label{sec:3}

 \subsection{Experimental setup}

    We evaluated the proposed models using Google's Speech Commands Dataset version 1 \cite{warden2017launching}.
    The dataset consists of 64721 one-second long recordings of 30 words by thousands of different speakers, as well as background noise samples such as pink noise, white noise, and human-made sounds. Among the 30 words, 10 words (including ``yes'', ``no'', ``up'', ``down'', ``left'', ``right'', ``on'', ``off'', ``stop'', ``go'') were used as keywords, and the rest 20 words were used as fillers which were labeled as ``unknown''.

   We followed the way in \cite{tang2018deep} to add noise and random shift to each segment. Then, we extracted 40 dimensional Mel-frequency cepstrum coefficient features from each frame with a frame length of 25ms and a frame shift of 10ms.
   We used the stochastic gradient descent with a momentum of 0.9 as the optimizer of the proposed networks, and added the $L_{2}$ weight decay of $10^{-3}$ as the regularization.
   The batch size was set to 100. All models were trained from scratch for roughly 30000 steps. The initial learning rate was set to 0.1, and multiplied by 0.1 for every 10000 steps.
   The network was evaluated on the validation set for every 1000 steps. The model that achieved the highest accuracy on the validation set was used as the final model. We ran all experiments for five independent times with different random seeds, and reported the average performance.

 \subsection{Results of the first experiment}

   In the first experiment, we followed the experimental setup of \cite{tang2018deep}. Specifically, we additionally added some silent segments which are random noise only. We assigned the silent segments a keyword ``silence''. We randomly selected a number of segments from the keyword ``unknown'', which keeps the ratio of ``silence'' and ``unknown'' to about 10\% of the total segments.
   According to the SHA1-hashed name, the audio files was split to three parts for training, validation, and test, which contain roughly 22000, 2700, and 2700 segments, respectively.

   We compared the proposed DS-ResNet with ResNet \cite{tang2018deep}, TC-ResNet \cite{choi2019temporal}, and DS-CNN \cite{zhang2017hello}. The three models of ResNets are denoted as \texttt{res15}, \texttt{res15-narrow}, and \texttt{res8-narrow}.
   The three models of TC-ResNets are denoted as \texttt{TC-ResNet14-1.5}, \texttt{TC-ResNet14}, and \texttt{TC-ResNet8}. The three models of DS-CNNs are denoted as \texttt{DS-CNN(L)}, \texttt{DS-CNN(M)}, and \texttt{DS-CNN(S)}. See \cite{tang2018deep,choi2019temporal,zhang2017hello} for the description of the aforementioned comparison models.
   All comparison methods followed the same settings as in \cite{tang2018deep,choi2019temporal,zhang2017hello}. 

   \begin{figure}[t]
     \centering
     \includegraphics[scale=0.39]{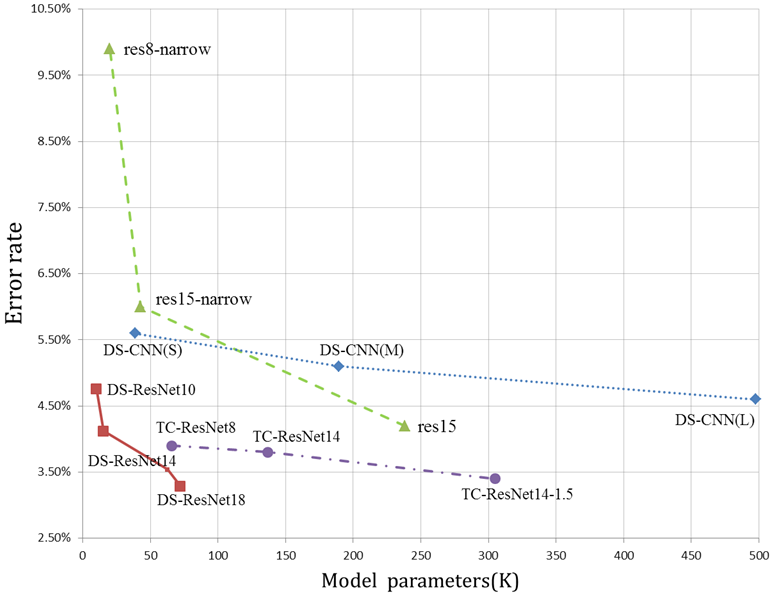}
     \caption{Performance curves of the comparison methods.}
     \label{fig:result_res}
   \end{figure}

   \begin{table}[t]
      \caption{Comparison between DS-ResNet and ResNet. The number after ``$\pm$'' is the 95\% confidence interval.}
      \label{table:result_res}
      \centering
      \scalebox{0.83}{
      \begin{tabular}{l@{}r c r c}
        \toprule
        \multicolumn{2}{l}{\textbf{}} &
        \multicolumn{1}{c}{\textbf{Error rate}}&
        \multicolumn{1}{c}{\textbf{\#Parameters}} &
        \multicolumn{1}{c}{\textbf{\#Multiplies}}  \\
        \midrule
        \texttt{res15}*\cite{tang2018deep} & & $4.20\%\pm 0.484$ & 238K & 894M
        \\
        \texttt{res15-narrow}*\cite{tang2018deep} & & $6.0\%\pm 0.516$  & 42.6K & 160M
        \\
        \texttt{res8-narrow}*\cite{tang2018deep} & & $9.90\%\pm 0.976$  & 19.9K & 5.65M
        \\
        \texttt{DS-ResNet18} & & $3.29\%\pm0.195 $  & 72K & 285M        \\
        \texttt{DS-ResNet14} & & $4.12\%\pm0.178 $  & 15.2K & 15.7M
        \\
        \texttt{DS-ResNet10} & & $4.76\%\pm0.365 $  & 10K & 5.8M             \\
        \bottomrule
      \end{tabular}
      }
   \end{table}

   Figure \ref{fig:result_res} shows the comparison results in terms of the error rate (in the vertical ordinate) and model parameters (in the horizontal ordinate). From the figure, we see that the proposed DS-ResNet yields better performance curve than the comparison methods.

   Table \ref{table:result_res} lists the comparison results in terms of the error rate, model parameters, and the number of multiplication operations per inference pass. From the table, we see that \texttt{DS-ResNet18} achieves a relative 21.7\% error rate reduction over \texttt{res15}, with its number of parameters being only 1/3 of that of the latter. When the number of the model parameters is roughly the same, \texttt{DS-ResNet14} achieves a relative 58.4\% error rate reduction over \texttt{res8-narrow}. At last, \texttt{DS-ResNet10} reaches an error rate of 4.76\% with only 10K parameters.

  \subsubsection{Effect of the squeeze-and-excitation block}

   To investigate the effect of the squeeze-and-excitation block, we proposed three additional variants of \texttt{DS-ResNet18}. The first one, named \texttt{DS-ResNet18-n}, does not use the squeeze-and-excitation block. The second one, named \texttt{DS-ResNet18-d}, added the squeeze-and-excitation block after each depthwise convolution layer. The third one, named \texttt{DS-ResNet18-p}, added the squeeze-and-excitation block after each pointwise convolution layer.

Table \ref{table:result_se} lists the effect of the squeeze-and-excitation block on performance.
   From the table, we see that \texttt{DS-ResNet18} outperforms \texttt{DS-ResNet18-n}, which demonstrates the effectiveness of the block.
   However, adding more squeeze-and-excitation blocks, as the \texttt{DS-ResNet18-d} and \texttt{DS-ResNet18-p} did, does not lead to improved performance.
  \begin{table}[t]
      \caption{Effect of the squeeze-and-excitation block with different settings.}
      \label{table:result_se}
      \centering
      \scalebox{0.83}{
      \begin{tabular}{l@{}r r r}
        \toprule
        \multicolumn{2}{l}{\textbf{}} &
        \multicolumn{1}{l}{\textbf{Error rate}}&
        \multicolumn{1}{l}{\textbf{\#Parameters}}
        \\
        \midrule
        \texttt{DS-ResNet18} & & $3.29\%\pm0.195 $  & 72K
        \\
        \texttt{DS-ResNet18-n} & & $3.45\%\pm 0.152$ & 71.4K
        \\
        \texttt{DS-ResNet18-d} & & $3.54\%\pm 0.191$  & 79.6K
        \\
        \texttt{DS-ResNet18-p} & & $3.67\%\pm0.147 $  & 79.6K               \\
        \bottomrule
      \end{tabular}
      }
   \end{table}

  \subsection{Results of the second experiment}

To further investigate the effectiveness of the proposed method, we used the standard configuration of the dataset \cite{ZengEffective,bai2019time}, where we used 51088 utterances for training, 6798 utterances for validation, and 6835 utterances for testing. We compared with the \texttt{DenesNet-BiLSTM} \cite{ZengEffective} and \texttt{tdnn-swsa} \cite{bai2019time}. All comparison methods followed the same settings as in \cite{ZengEffective,bai2019time}.

    \begin{table}[t]
      \caption{Comparison between DS-ResNet, DenesNet-BiLSTM, and tdnn-swsa.}
      \label{table:result_other}
      \centering
      \scalebox{0.83}{
      \begin{tabular}{l@{}c c r}
        \toprule
        \multicolumn{2}{l}{\textbf{}} &
        \multicolumn{1}{l}{\textbf{Error rate}}&
        \multicolumn{1}{l}{\textbf{\#Parameters}}
        \\
        \midrule
        \texttt{DenesNet-BiLSTM}*\cite{ZengEffective} & & $2.5\%$  & 250K
        \\
        \texttt{tdnn-swsa}*\cite{bai2019time} & & $4.19\%\pm 0.191$ & 12K
        \\
        \texttt{DS-ResNet18} & & $2.32\%\pm0.109 $  & 72K        \\
        \texttt{DS-ResNet14} & & $2.84\%\pm0.257 $  & 15.2K               \\
        \texttt{DS-ResNet10} & & $3.97\%\pm0.154 $  & 10K               \\
        \bottomrule
      \end{tabular}
      }
   \end{table}

  Table \ref{table:result_other} lists the comparison results. From the table, we see that \texttt{DS-ResNet10} is competitive with \texttt{tdnn-swsa} in the low-resource condition. If we slightly relaxed the restrictions on the model size, \texttt{DS-ResNet14} achieves a relative 32.2\% error rate reduction over \texttt{tdnn-swsa}. Finally, \texttt{DS-ResNet18} achieves similar performance with \texttt{DenesNet-BiLSTM}, with the number of parameters being only 1/3 of the latter.

\section{Conclusions}\label{sec:4}

In this paper, we have proposed the depthwise separable convolution based ResNet for the small-footprint keyword spotting problem, which contains two novel components. The first component concatenates the depthwise separable convolution with ResNet. This concatenation significantly reduces the number of parameters without suffering performance degradation. The second component applies the squeeze-and-excitation block to the output of the first convolution layer, which is able to further improve the performance without increasing the number of parameters dramatically. We have compared DS-ResNet with 5 referenced methods on two settings of the public available Google Speech Commands dataset. Experimental results show that the proposed DS-ResNet achieves the state-of-the-art performance in various experimental settings.
%

%

\bibliographystyle{IEEEtran}
\normalem
\bibliography{mybib}


\end{document}